\documentclass[prl,twocolumn,showpacs,amsmath,amssymb,superscriptaddress]{revtex4-1}

\usepackage{amsmath,amssymb,amsfonts,bm,color,graphicx,tabularx,physics}

\usepackage[unicode=true,colorlinks=true]{hyperref}

\usepackage[etex=true,export]{adjustbox}

\hypersetup{linkcolor=blue,citecolor=blue,urlcolor=blue}
\usepackage{diagbox}

\newcommand{\beq}{\begin{equation}}
	\newcommand{\eeq}{\end{equation}}
\newcommand{\beql}{\begin{equation*}}
	\newcommand{\eeql}{\end{equation*}}
\newcommand{\beqn}{\begin{eqnarray}}
	\newcommand{\eeqn}{\end{eqnarray}}

\begin{document}
	
	\title{Anisotropic and tunable vortex topology in multiband iron-based superconductors}

         \author{Si-Qi Yu}
          \thanks{These authors contributed equally to this work.}
	\author{Wei Cheng}
          \thanks{These authors contributed equally to this work.}
	\affiliation{School of Physics and Wuhan National High Magnetic Field Center,  Huazhong University of Science and Technology, Wuhan, 430074, China}
	\affiliation{ Institute for Quantum Science and Engineering and Hubei Key Laboratory of Gravitation and Quantum Physics,  Huazhong University of Science and Technology, Wuhan, 430074, China}
	\author{Chuang Li}
         \affiliation{School of Physics and Wuhan National High Magnetic Field Center,  Huazhong University of Science and Technology, Wuhan, 430074, China}
	\author{Xiao-Hong Pan}
       \affiliation{Department of Physics, College of Physics, Optoelectronic Engineering, Jinan University, Guangzhou 510632, China}
        \affiliation{School of Physics and Wuhan National High Magnetic Field Center,  Huazhong University of Science and Technology, Wuhan, 430074, China}
         \author{Gang Xu}
\email{gangxu@hust.edu.cn}
        \affiliation{School of Physics and Wuhan National High Magnetic Field Center,  Huazhong University of Science and Technology, Wuhan, 430074, China}
	\affiliation{ Institute for Quantum Science and Engineering and Hubei Key Laboratory of Gravitation and Quantum Physics,  Huazhong University of Science and Technology, Wuhan, 430074, China}
\affiliation{Wuhan Institute of Quantum Technology, Wuhan 430206, China}
     
	\author{Fu-Chun Zhang}
	\email{fuchun@ucas.ac.cn}
	\affiliation{Kavli Institute for Theoretical Sciences, University of Chinese Academy of Sciences, Beijing 100190, China}
      \affiliation{Hefei National Laboratory, Hefei 230088, China}
	
	\author{Xin Liu}
	\email{phyliuxin@hust.edu.cn}
   \affiliation{School of Physics and Wuhan National High Magnetic Field Center,  Huazhong University of Science and Technology, Wuhan, 430074, China}
\affiliation{Tsung-Dao Lee Institute and School of Physics and Astronomy, Shanghai Jiao Tong University, Shanghai 201210, China }
\affiliation{ Institute for Quantum Science and Engineering and Hubei Key Laboratory of Gravitation and Quantum Physics,  Huazhong University of Science and Technology ,Wuhan, 430074, China}
\affiliation{Hefei National Laboratory, Hefei 230088, China}
\affiliation{Wuhan Institute of Quantum Technology, Wuhan 430206, China}
	\date{\today}
	\begin{abstract}
		Building on the multiband nature of iron-based superconductors (FeSCs), we have uncovered pronounced anisotropy in Majorana vortex topology arising from the interaction between vortex orientation and multiple electronic topologies. This anisotropy manifests in two distinct vortex configurations: the z-vortex and x-vortex, oriented perpendicular and parallel to the Dirac axis (z-axis for FeSCs), respectively. The x-vortex exhibits a unique duality, displaying two distinct topological phase diagrams. One is strikingly simple, comprising only trivial and topological superconducting phases, and remains resilient to multiband entanglement. The other mirrors the z-vortex's complex diagram, featuring alternating trivial, topological crystalline and topological superconducting phases. Crucially, the former is exclusive to the x-vortex and supports unpaired Majorana vortices across a wide parameter range, even with Dirac nodes in electronic bands. Notably, uniaxial strain can modulate these x-vortex phases, enabling the x-vortex to support both stable Majorana vortices and rich exotic physics in a controllable manner. Moreover, we propose that the x-vortex offers promising advantages for developing iron-based superconducting quantum devices. Our findings introduce a novel paradigm in vortex topology within multiband superconducting systems, highlighting the x-vortex as a promising platform for exploring Majorana physics and advancing iron-based superconducting quantum technology.  
	\end{abstract}
	\maketitle
	
	{\it{Introduction}}--  The multiband nature of iron-based superconductors (FeSCs) endows them with not only unique high-temperature superconductivity such as $s_{\pm}$ pairing \cite{christianson_unconventional_2008,mazin_unconventional_2008,Lee2008,Cvetkovic2009,hanaguri_unconventional_2010,wang2010,Pietosa2012,Borisenko2012,Lin2013,Chang2014,Chen2014,Miao2015,Nag2016,Zhang2019,chen_direct_2019,Yin2019,wu_nodal_2024s+-} and fractional vortices \cite{iguchi_superconducting_2023,zheng_direct_nodate,zhou_observation_nodate} but also rich topological band structures \cite{Nekrasov2008,Singh2008,hao_topological_2014,Ye2014,wang_topological_2015,Borisenko2016,wu_topological_2016,hao_topological_2019,Chen2021,Qin2022,Ma2022,Yang2023}. The coexistence of superconductivity and topological bands in principle replicates the Fu-Kane paradigm \cite{Fu2008} of topological SCs in a single material system, positioning them as promising platforms for realizing Majorana zero modes (MZMs) \cite{XGang-2016-PRL,zhang_observation_2018,Qin2019a,Hu2024,Cao2024}, the simplest non-Abelian anyons with immense potential for topological quantum computation (TQC) \cite{Moore-1991-NPB,Ivanov-2001-PRL,Kitaev-2003-AnnPhys,DasSarma-2006-PRB,DasSarma-2008-RMP,HXiao-2012-EPL, HXiao-2014-STAM, HXiao-2014-PRB}. Experimental evidence, including observations of zero-bias conductance peaks and quantized Caroli states in various FeSCs \cite{Yin2015,zhang_observation_2018,DHong-2018-Sci,yim_discovery_2018,WHaiHu-2018-NatC,FDongLai-2018-PRX,FDongLai-2019-CPL,Hanaguri-2019-NatM,KLingYuan-2019-NatPhys,ZPeng-2019-NatPhys,DHong-2020-NatCom, DHong-2020-Sci,Zhang2021}, supports this expectation. 
	
	However, the multiband nature of FeSCs presents a double-edged sword: while it engenders rich phenomena, it also complicates the underlying physics \cite{Hu2022}. The coexistence of TI band and Dirac nodes can lead to the coexistence of  gapless and Majorana vortices \cite{konig_crystalline-symmetry-protected_2019,Qin2019,konig_crystalline-symmetry-protected_2019,Zhang2023a} which suppresses the MZMs' stabablity, reflecting in the lower ratio and even the lack of MZMs evidence in FeSeTe \cite{DHong-2018-Sci,Wu2021,chiu2020scalable} and LiFeAs \cite{Hanaguri2012} respectively. The strain-induced $C_{4z}$ rotational symmetry breaking has yielded some Majorana evidence \cite{Kong2021,fan_observation_2021,Li2022a, Liu2022c,Li2023,hu_observation_nodate}, albeit in a limited parameter range. These findings suggest that the multiband nature of FeSCs may hinder the realization and detection of Majorana vortices. Notably, previous studies have predominantly focused on z-vortices, where the magnetic field aligns with the Dirac nodes. Meanwhile, the development of strategies for Majorana vortex braiding in FeSCs remains an open challenge, with recent research on iron-based nanowires \cite{chang_growth_2014,tao_synthesis_2018,liu_characterization_2022} offering promising directions. To the best of our knowledge, so far the grown iron-based superconducting nanowire is in the a-b plane.   
	
	\begin{figure}
		\centering
		\includegraphics[width=1\columnwidth]{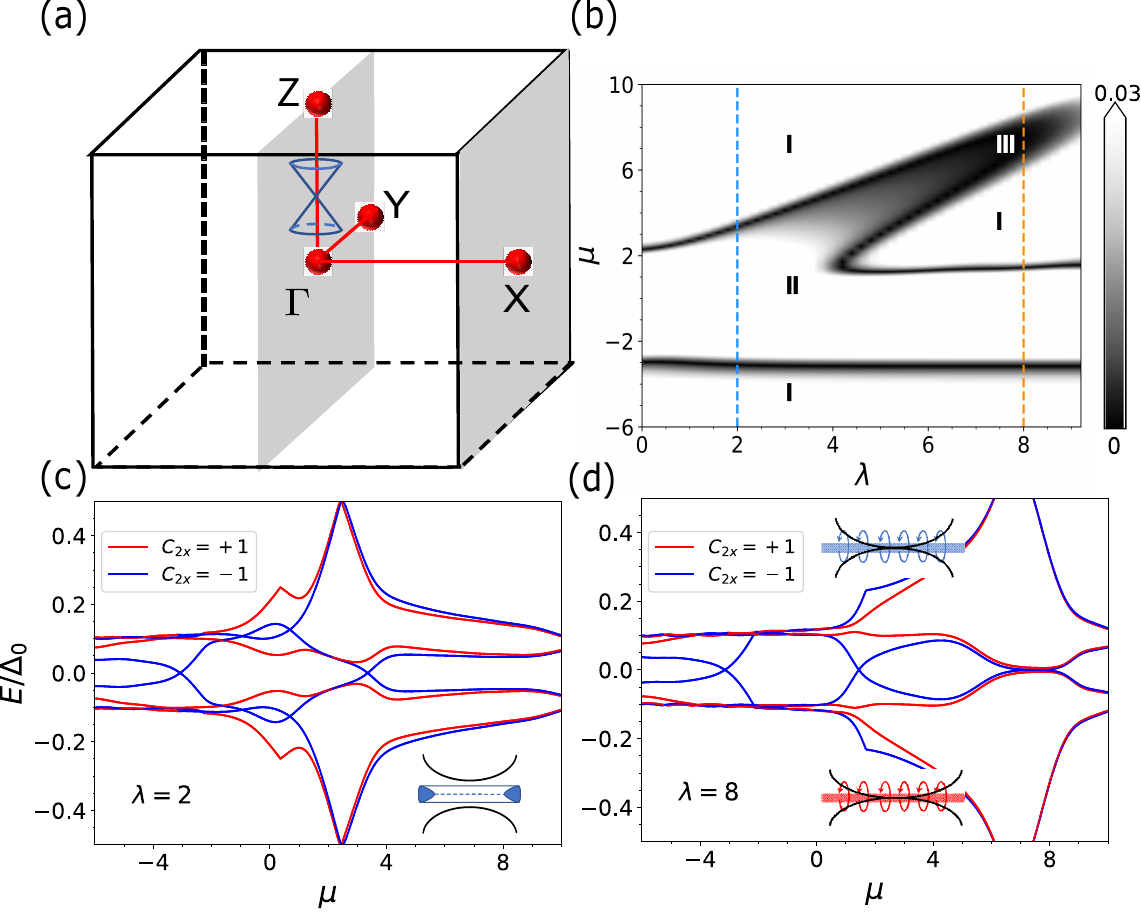}
		\caption{(a) Schematic representation of the Brillouin zone for iron-based superconductors. (b) Phase diagram of the x-vortex in the $\mu$ and SOC parameter space, while maintaining $C_{4z}$ symmetry. The colorbar represents the absolute value of energy closest to the Fermi surface in $k_x = 0$ plane, renormalized by the superconducting gap $\Delta_0$. Panels (c) and (d) show the $E-\mu$ spectrum near the Fermi surface at $k_x = 0$, corresponding to line cuts in panel (b). The red (blue) curves represent the eigenvalues of the states in $C_{2x} = +1 (-1)$ subspace.}
		\label{Figure_1}
	\end{figure}

	In this work, we demonstrate that the multiband nature of FeSCs introduces significant anisotropy into the vortex phase diagram. We explore vortex physics under a magnetic field applied in the y-z plane, focusing on the vortex along the x-direction, referred to as the x-vortex. Our analysis reveals two distinct phase diagrams. One closely resembles the well-studied z-vortex phase diagram, exhibiting the coexistence of gapless and Majorana vortices with $C_{4z}$ symmetry, as well as two Majorana vortices when $C_{4z}$ symmetry is broken. In this case, x(z)-vortices are protected by $C_{2x(z)}$ symmetry and are topologically equivalent to the topological crystalline superconducting phase \cite{liu_signatures_2024,wan_large-scale_2024}. The second phase diagram, unique to x-vortices, consists only of conventional and topological superconducting phases, supporting an unpaired Majorana vortex over a wide range of chemical potential. This phase diagram is exclusive to the x-vortex and demonstrates that FeSCs can support stable MZMs immune to multiband entanglement. Notably, these phase transitions are closely tied to the transition from type I to type II Dirac nodes under $C_{4z}$ rotational symmetry, as well as to the controllable mirror Chern numbers induced by uniaxial strain at the $k_x = 0$ mirror-invariant plane. Furthermore, we propose that x-vortices offer significant advantages for developing iron-based superconducting quantum devices. Our findings introduce a new paradigm for Majorana vortex phase transitions in multiband superconducting systems, highlighting the x-vortex configuration as a promising platform for advancing Majorana physics in FeSCs and realizing novel quantum devices.

	{\it Theoretical model}--
	We employ the Bogoliubov-de Gennes (BdG) Hamiltonian to model the superconducting system, expressed as
	\begin{align} \label{eq:FeSC-Hmtn}
		H_\text{S}= \begin{pmatrix}
			H_\text{e}(\bm{r}) -\mu & \hat{\Delta}(\bm{r}) \\
			\hat{\Delta}^{\dagger}(\bm{r}) & -H_\text{e}^*(\bm{r}) +\mu
		\end{pmatrix} \ ,
	\end{align}
	under the Nambu basis $\{\Psi_r^\dagger,\Psi_r^T\}$, with $H_{\rm e}$ the electronic Hamiltonian, $\mu$ the chemical potential, and $\hat{\Delta}(\bm{r})$ the superconducting gap function. To capture the multiband characteristics of FeSCs, we express $H_{\rm e}(\bm r)$ in a six-band basis, consisting of the states: $(|z\uparrow\rangle, |z\downarrow\rangle, |-i (x + iy) z\downarrow\rangle, |-i (x - iy) z\uparrow\rangle, |-i (x + iy) z\uparrow\rangle, |i (x - iy) z\downarrow\rangle)$. The electronic Hamiltonian in momentum space is then given by
	\begin{align}\label{Ham-1}
		H_{\rm e}(\mathbf{k}) =& \epsilon(\mathbf{k}) +\begin{pmatrix}
			M(\mathbf{k}) & T_{\rm T}(\mathbf{k}) & T_{\rm D}(\mathbf{k})  \\
			T_{\rm T}^\dagger(\mathbf{k}) & -M(\mathbf{k})& T_{\rm F}(\mathbf{k}) \\
			T_{\rm D}^\dagger(\mathbf{k})  & T_{\rm F}^\dagger(\mathbf{k}) & -M(\mathbf{k})+\lambda 
		\end{pmatrix},
	\end{align}
	where 
	\begin{align}\label{Ham-e}
		\epsilon(\mathbf{k}) =& C +2D(2-\cos k_x - \cos k_y) +2D_3(1-\cos k_z), \nonumber \\
		M(\mathbf{k}) =& M_0 +2B(2-\cos k_x -\cos k_y) +2B_3(1-\cos k_z), \nonumber \\
		T_{\rm T}(\mathbf{k}) =& A(\hat{s}_x\sin k_x +\hat{s}_y\sin k_y) +A_3\hat{s}_z\sin k_z, \nonumber \\ 
		T_{\rm D}(\mathbf{k}) =& A(\hat{s}_z\sin k_x+i\sin k_y) + A'_3 \hat{s}_x \sin k_z , \nonumber \\
		T_{\rm F}(\mathbf{k}) =& i[\alpha\hat{s}_x \sin k_x \sin k_y -\beta\hat{s}_y(\cos k_x -\cos k_y)],
	\end{align}
	with $s$ the Pauli matrix acting on the spin space and $\lambda$ the spin-orbit coupling (SOC) strength. Through this work, with $A_3 = 1$, we take $C = 4$, $D = 0$, $D_3 = -2$, $M_0 = 6$, $B = 16$, $B_3 = -3$, $A = 10$ to be consistent with the experiment \cite{ZPeng-2019-NatPhys}. Note that $2A_3$ corresponds to the size of the TI gap, and we define $\lambda \gg 2A_3 $ as the high SOC regime to characterize the SOC strength. When considering an x-vortex, the superconducting gap function takes the form  $\hat{\Delta}(\bm{r})=i\Delta(r)\hat{s}_y$, where $\Delta(r)=\Delta_0 \tanh(r/\xi) e^{i\phi}$, $r=\sqrt{y^2+z^2}$, $\xi$ is the superconducting coherence length, and $\cos(\phi) = y/r$. We take $\Delta_0=2$ for calculation convenience \cite{supp}.  
	
	Note that $A'_3=0$ results in the Dirac node at $\Gamma-Z$ line (Fig.~\ref{Figure_1}(a)) which is protected by $C_{4z}$ symmetry \cite{Wang2015f}. The coexistence of the Dirac nodes and the TI bands complicates the visualization of the vortex topology from the two-dimensional topological surface states. Alternatively, the vortex line can be viewed as a quasi-1D system that belongs to class D in the Altland-Zirnbauer classification \cite{Vishwanath-2011-PRL}. From this perspective, the existence of MZMs requires two conditions: 
	1. The sign change of $\mathcal{Z}_2$ index at the mirror symmetry planes for $k=0$ and $k=\pi$ \cite{Vishwanath-2011-PRL}, referred to as the vortex-phase-transition-plane (VPTP), with $k$ being the momentum along the magnetic field direction. 
	2. A fully gapped quasi-one-dimensional system. In prior studies of z-vortices, the Dirac nodes lie outside the VPTP and thus only affect the second condition. In contrast, for the x-vortex, the Dirac nodes reside within the VPTP (Fig.~\ref{Figure_1}(a)), affecting both conditions simultaneously. This difference also applies to $A'_3 \neq 0$ and is crucial for understanding the anisotropy of the Majorana vortex phase in multiband FeSCs, as we will illustrate below. 
	
	\begin{figure}
		\centering
		\includegraphics[width=1\columnwidth]{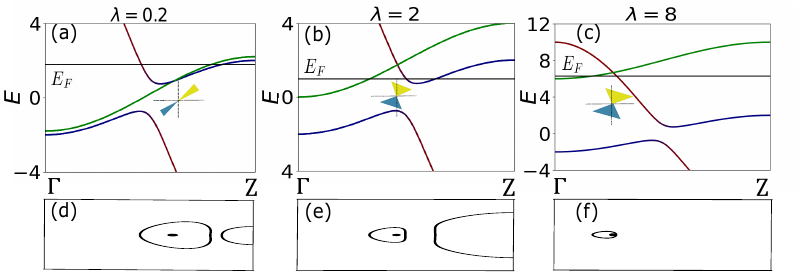}
		\caption{(a)-(c) The electronic band structure and Dirac points evolve with changes in SOC strength $\lambda$ when the system exhibits $C_{4z}$ symmetry. The colors represent the contributions from different orbitals, with red, green, and blue denoting the $p_z$, $d_{xz}$, and $d_{yz}$ orbitals, respectively. The tilting of the Dirac cones is characterized by these changes. The corresponding Fermi surface near the Dirac point  are depicted in (d)-(f). }
		\label{Dirac_nodel}
	\end{figure}
	
	{\it Vortex topology with Dirac nodes}-- We first examine the vortex states when the electronic bands respect C$_{4z}$ symmetry with $A'_3 = 0$, resulting in Dirac nodes \cite{Wang2015f}.  The phase diagram in the $\lambda-\mu$ parameter space (Fig.~\ref{Figure_1}(b)) displays the lowest eigenvalue of $H_{S}(k_x=0,y,z)$. In the high SOC regime ($\lambda \gg A_3$), we observe gapless and Majorana vortex states when the chemical potential is near the Dirac node and the TI band gap, respectively. The vortex phase transition exhibits standard multiband features (around the orange dashed line in Fig.~\ref{Figure_1}(b)), similar to those in the z-vortex. Notably, as the SOC strength decreases to $\lambda ~ A_3$ o smaller, the gapless vortex merges into the gapped Majorana vortex, with the phase diagram (around the blue dashed line in Fig.~\ref{Figure_1}(b)) behaving similarly to the SC/TI hybrid system \cite{Fu2008,Vishwanath-2011-PRL}. To further elucidate the vortex phases, we compare the energy spectrum versus chemical potential from low (Fig.~\ref{Figure_1}(c)) to high SOC (Fig.~\ref{Figure_1}(d)) strengths. The vortex system's $C_{2x}$ rotational symmetry allows the classification of its low-energy spectrum into $C_{2x} = \pm 1$ subspaces. In the high SOC case, around the Dirac node, each subspace supports a gapless vortex state, while around the TI band gap, only the $C_{2x}=-1$ subspace supports a Majorana vortex. This is consistent with viewing the multi-band system as a combination of independent Dirac and TI bands. In the low SOC case, the $C_{2x}=-1$ subspace supports one unpaired Majorana vortex, while the $C_{2x}=+1$ subspace supports a trivial gapped vortex. These results indicate that the x-vortex in iron-based superconductors with low SOC strength naturally supports unpaired MZMs despite the persistent coexistence of Dirac and TI bands.
	
	\begin{figure*}
		\centering
		\includegraphics[width=2\columnwidth]{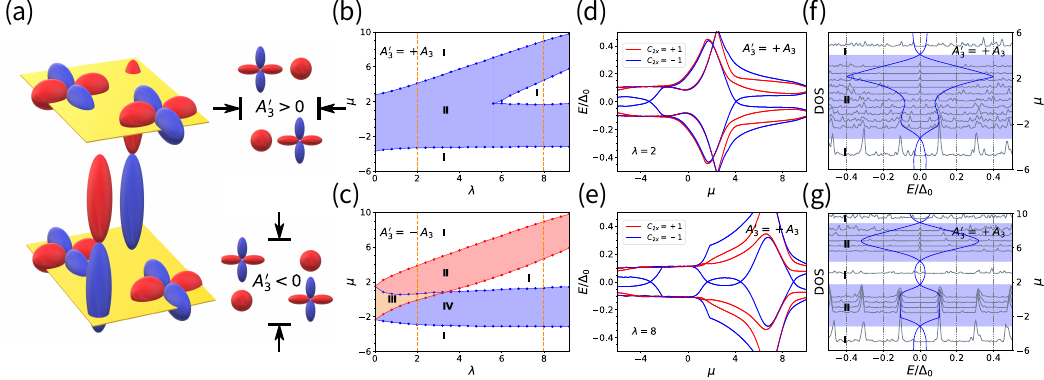}
		\caption{(a) Schematic diagram of the orbital. The right side of (a) illustrates lattice deformation along the $a$- or $b$-axis under applied strain. With the breaking of $C_{4z}$ symmetry, (b) presents the phase diagram in the $\mu\text{-}\lambda$ parameter space for $A_3^\prime > 0$, while (c) shows the phase diagram for $A_3^\prime < 0$. (d) and (e) display the energy spectrum as a function of $\mu$ at $k_x = 0$, corresponding to the line cuts in (b) and (c), respectively. (f) and (g) show the eigenvalues closest to the Fermi level (blue curves) with the periodic boundary condition and the DOS spectra (gray curves) with the open boundary condition for the parameters used in (d) and (e). The lattice size is up to $N_x=120,N_y=60,N_z=30$ in the calculation of DOS. }
		\label{Fig_2}
	\end{figure*}
	
	The vortex topological phase transition is related to having $\pi$ Berry phase around the electronic Fermi surface on the VPTP \cite{Hosur2011,supp}. The electronic Hamiltonian in Eq.~\eqref{Ham-e} has mirror-x symmetry \cite{supp}. Therefore, the electronic Hamiltonian on the VPTP, the mirror-x invariant plane, commutes with the mirror operator $M_x = \text{diag}[i, -i, i] \otimes \hat{s}_x$ and can be block diagonalized into the $M_x = \pm i$ subspace. The two subspaces are related by time-reversal symmetry, and hence we analyze only the $M_x = +i$ subspace for simplicity. In this case, the lowest band is always gapped from the other two bands and carries a Chern number of $-1$ in the $M_x = +i$ subspace \cite{supp}. Consequently, there is always a Berry phase of $\pi$ around certain Fermi surfaces in the lowest band, giving the phase transition in the lower chemical potential region (Fig.~\ref{Figure_1}(b)). With this in mind, let us focus on the Dirac nodes formed by the top two bands. We found a transition from type I to type II Dirac points (Fig.~\ref{Dirac_nodel}) when Lowering SOC strength $\lambda$ from $\lambda > A_3$ to $\lambda < A_3$. The Fermi surface of type II Dirac points is non-closed, rendering the Berry phase ill-defined for type II Dirac points, whereas it is quantized to $\pm \pi$ for type I Dirac points. The absence of a $\pi$ Berry phase for type II Dirac nodes impedes the formation of gapless vortices and inhibits the topological vortex phase transition around the Dirac point. This behavior is consistent with the vortex phase of a single Dirac metallic band \cite{YZhongbo-2020-PRL,supp}. However, since the topological phase transition already occurs once in the TI valence band (bottom black line in Fig.~\ref{Figure_1}), a second phase transition must be required \cite{supp}. Our numerical results show that this transition occurs above the Dirac point. Thus, the phase diagram with the type II Dirac nodes resembles a conventional superconducting topological insulator \cite{Fu2008, Hosur2011}. This suggests that a lower SOC strength, leading to the type II Dirac semimetal phase, is more favorable for realizing MZMs in vortices. This is manifestly due to the multiband nature of iron-based superconductors and is in sharp contrast to the single Dirac band case \cite{YZhongbo-2020-PRL}.

	{\it Vortex topology without C$_{4z}$ symmetry}-- Recent research has shown that breaking $C_{4z}$ symmetry is crucial for realizing z-vortex topology in iron-based superconductors. Building on this insight, we extend our investigation to the x-vortex topology by incorporating strain-induced $C_{4z}$ symmetry breaking. Notably, $A_3'=t'_x \cos k_x - t'_y \cos k_y$ arise from the SOC-assisted inter-layer hopping, with $t'_x$ and $t'_y$ the hopping strength \cite{supp}, between the Fe's $d$-orbital and As's or Se's $p_z$-orbital (Fig.~\ref{Fig_2}(a)). Breaking the $C_{4z}$ symmetry by allowing $t'_x \neq t'_y$ results in a gap opening at the Dirac nodes, leading to two TI band structures in the system. This can be achieved by applying compressive (tensile) strain along the x-direction increases (decreases) the hopping strength along x relative to y \cite{Li2022a, Liu2022c,Li2023,hu_observation_nodate}. Consequently, the term $A'_{3} = t'_{3x}\cos k_x - t'_{3y}\cos k_y$, with $t'_{3x} > t'_{3y}$ (or $t'_{3x} < t'_{3y}$), opens a Dirac gap and yields $A_{3}' > 0$ (or $A_{3}' < 0$) around the $\Gamma-Z$ line. This modification also opens a gap in what was previously a gapless vortex region in the high SOC regime of the x-vortex (Fig.~\ref{Fig_2}(b,c)), similar to the z-vortex case. However, the x-vortex exhibits a unique characteristic: it possesses two distinct phase diagrams, depending on the sign of $A_{3}'$, whereas the z-vortex has only a single phase diagram. When $A_{3}'<0$, the phase diagram closely mirrors that of the z-vortex across all regimes. In the high SOC regime, two distinct TSC phases emerge (Fig.~\ref{Fig_2}(c)), each residing in the $C_{2x} = \pm 1$ subspace \cite{supp}. In the low SOC regime, these two TSC phases coexist, stabilized by the $C_{2x}$ rotational symmetry (region III in Fig.~\ref{Fig_2}(c)). For $A_{3}'>0$ in the high SOC regime, both TSC phases (Fig.~\ref{Fig_2}(e)) reside in the $C_{2x}=-1$ subspace. As the SOC strength decreases, the two TSC phases merge into a single phase rather than coexisting (Fig.~\ref{Fig_2}(d)). The density of states calculations (Fig.~\ref{Fig_2}(f,g)) show how the zero-bias peak evolves with changes in chemical potential, consistent with the topological phase diagram in Fig.~\ref{Fig_2}(b). While the behavior for $A_{3}'<0$ is similar to that observed in the z-vortex, the $A_{3}'>0$ case is unique to the x-vortex. Remarkably, in the low SOC limit, the x-vortex can resemble the simpler Fu–Kane Majorana paradigm. In this regime, robust MZMs emerge without being affected by the multiband complexity inherent in iron-based superconductors.
	
\begin{figure}
		\centering
		\includegraphics[width=1.1\columnwidth]{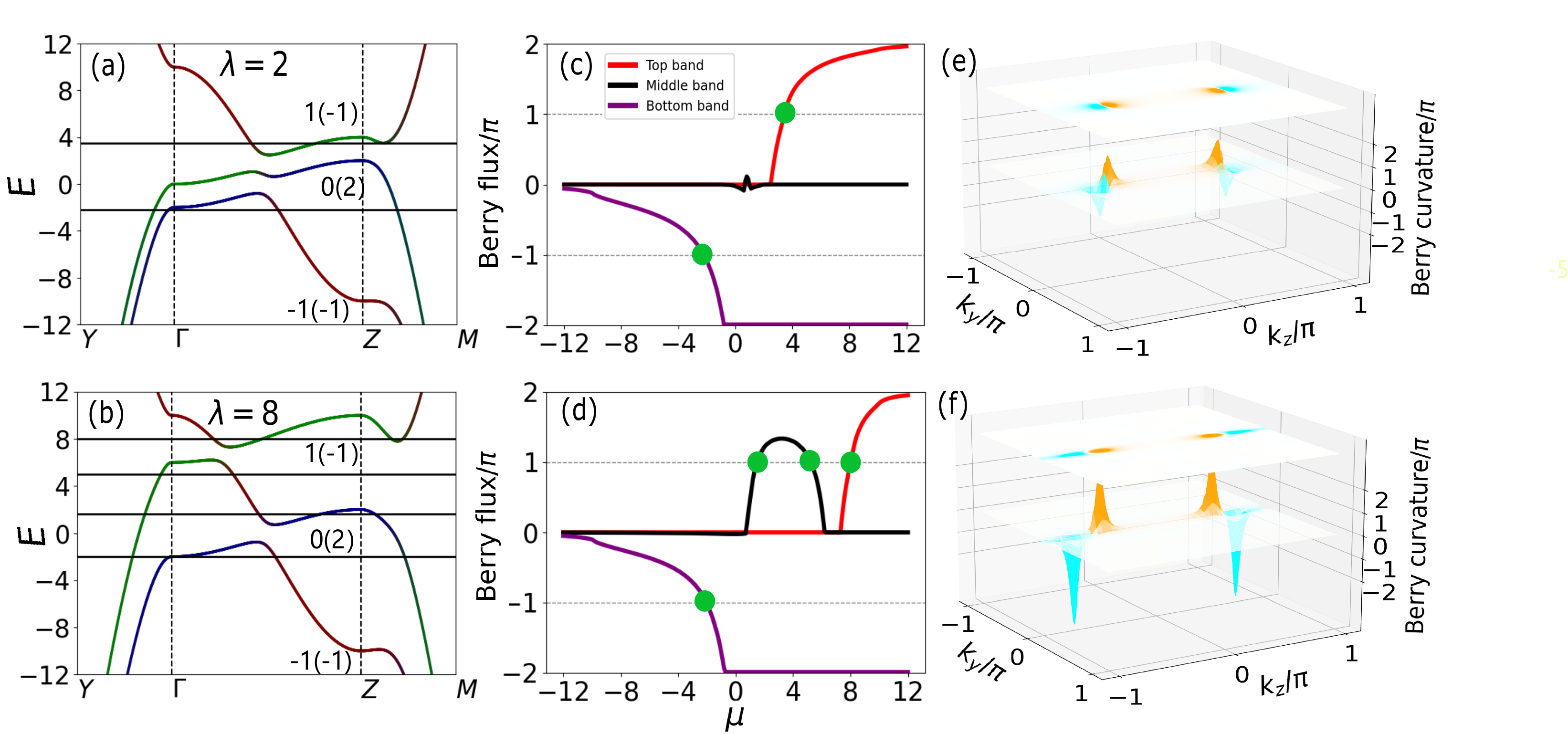}
		\caption{(a) and (b) show the electronic band structures for spin-orbit coupling (SOC) strengths $\lambda = 2$ and $8$, respectively. The colors represent the contributions from different orbitals, with red, green, and blue denoting the $p_z$, $d_{xz}$, and $d_{yz}$ orbitals, respectively. (c) and (d) illustrate the variation in the Berry phase across different bands as a function of chemical potential, with distinct colors corresponding to different bands when $A_{3}^{'}=A_{3}$. The green dots indicate the positions on the Fermi surface where the Berry phase accumulates a $\pi$ phase, marking the presence of a vortex-phase-transition. (e) and (f) display the Berry curvature distribution in $k$ -space for the middle band in (c) and (d). The orange and cyan colors denote positive and negative Berry curvature, respectively.}
		\label{C_4BR}
	\end{figure}

	To understand the $A_{3}'$-dependent phase diagram, we examine the band topology through the Mirror Chern numbers on the $\Gamma-Y-Z$ plane (Fig.~\ref{C_4BR} (a,b)). Notably, the lowest band is always gapped from others, regardless of whether $C_{4z}$ symmetry is broken. As we previously demonstrated, the lowest band’s Chern number remains fixed at $c_{3}=-1$, resulting in $c_{1}+c_{2}=1$. When $A_{3}'$ changes sign, each Dirac point contributes a variation of $+1$ or $-1$ to the Chern number. Due to inversion symmetry, all Dirac valleys contribute identically, causing the mirror Chern number of each of the top two bands to change by $\pm 2$ when $A_{3}'$ changes sign. We find that $A_{3}'>0$ and $A_{3}'<0$ yield $(c_{1},c_{2})=(1,0)$ and $(c_{1},c_{2})=(-1,2)$, respectively (Fig.~\ref{C_4BR}). Remarkably, the Berry curvature behaves very differently in these two cases. For $A_{3}'>0$, the Berry curvature of the middle band oscillates between positive and negative due to $c_{2}=0$ (Fig.~\ref{C_4BR} (e,f)). The magnitude of the Berry curvature is smaller in the low SOC regime (Fig.~\ref{C_4BR} (e,f)), attributable to the flatter second band (Fig.~\ref{C_4BR} (a,b)). The Berry phase at the Fermi level $\mu$ for each band can be calculated by integrating the Berry curvature from the band bottom to the Fermi surface. For the second band, the absolute value of the Berry phase on the Fermi surface is much smaller than $\pi$ in the low SOC regime (Fig.~\ref{C_4BR} (c)) due to the relatively flat band. In this case, phase transitions only occur when the chemical potential aligns with the top or bottom bands, and the vortex phase diagram resembles that of a superconducting topological insulator \cite{Fu2008}. Increasing the SOC strength enhances the Berry curvature amplitude, allowing the Berry phase on some Fermi surfaces to exceed $\pi$ (Fig.~\ref{C_4BR} (d,f)). Since the total Chern number is zero, the Berry phase of the Fermi surface crosses $\pi$ twice in this case, leading to two phase transitions for the chemical potential within the second band along the $\Gamma-Z$ line (Fig.~\ref{C_4BR} (d)). Consequently, the vortex phase diagram in the high SOC regime resembles that of the vortex along the c-direction. However, since these two phase transitions correspond to the same $\pi$ Berry phase, they belong to the same $C_{2x}$ section. For $A_{3}'<0$, the Berry phase on the Fermi surface of the second band continuously increases from $0$ to $4\pi$ due to $c_{2}=2$. As a result, there are always two phase transition points in the second band, regardless of the SOC strength \cite{supp}. Notably, these two phase transition points correspond to Berry phases of $\pi$ and $3\pi$, placing them in different $C_{2z}$ sections \cite{supp}. Therefore, the phase diagram for $A_{3}'<0$ qualitatively resembles that of the z-vortex.

	{\it Discussion--} The strong electronic correlations and multiorbital nature of these materials result in significant variations in their band structure, SOC strength, and chemical potential with changes in composition and doping \cite{yiObservationUniversalStrong2015, liuExperimentalObservationIncoherentcoherent2015, dayInfluenceSpinOrbitCoupling2018, Zhang2019c, dayThreedimensionalElectronicStructure2022a, maCorrelationcorrectedBandTopology2022a, kimOrbitalSelectiveMottTransition2024, liOrbitalIngredientsPersistent2024}. This variability complicates the observation of MZMs in the z-vortex phase, as evidenced by the trivial vortices in LiFeAs and the low ratio of Majorana vortices in FeSeTe. The complexity of the z-vortex phase diagrams likely contributes to these challenges. In contrast, the x-vortex phase diagram in a low SOC regime presents a simpler and more controllable environment, making it more favorable for observing vortex MZMs, even amid variations in material properties.  Moreover, FeSC quantum devices, such as FeSC nanowires or nanoribbons, are often oriented along the a-axis, aligning well with the required geometry for the x-vortex. The MZMs in the vortex core will appear at both ends of the nanowire or nanoribbon, enabling more controllable detection methods, such as correlation measurements of tunneling conductance. This nanowire setup is also compatible with current proposals for TQC in iron-Majorana platform \cite{Li2022}.

	{\it Conclusion}-- Our investigation reveals pronounced anisotropy in the vortex topology of iron-based superconductors. This anisotropy arises from the interaction between the vortex orientation and the multiband nature of FeSCs. Notably, despite the presence of multiband entanglement, the x-vortex phase diagram exhibits an extensive, strain-tunable topological region with a simple phase diagram. This finding highlights the potential of x-vortices for realizing stable Majorana vortices in multiband iron-based superconductors. Given the relevance of x-vortices to FeSC devices, our results pave the way to extend the Majorana vortex studies to quantum device which is the key for detecting non-Abelian braiding statistics and implementing TQC.

%	\bibliography{ref,new_yu,new_yu_scproperty}
	

\end{document}